# Liberal-Conservative Hierarchies of Intercoder Reliability Estimators

Yingjie Jay Zhao [a] (mc36547@um.edu.mo)

Guangchao Charles Feng [b] (fffchao@gmail.com)

Dianshi Moses Li [c] (yc37228@um.edu.mo)

Song Harris Ao [d] (harrisao@um.edu.mo)

Ming Milano Li [e] (yc17316@um.edu.mo)

Zhan Thor Tuo [c] (yc27203@connect.um.edu.mo)

Hui Huang [f] (hh@tenly.com)

Ke Deng [g] (kdeng@tsinghua.edu.cn)

Xinshu Zhao[*d] (xszhaoum@gmail.com)

[a]: Centre for Data Science, Institute of Collaborative Innovation, University of Macau, Taipa, Macao

[b]: Department of Interactive Media, School of Communication, Hong Kong Baptist University, Kowloon, Hong Kong

[c]: Centre for Empirical Legal Study, Faculty of Law, University of Macau, Taipa, Macao

[d]: Department of Communication, Faculty of Social Sciences, University of Macau, Taipa, Macao

[e]: Department of Government and Public Administration, Faculty of Social Sciences, University of Macau, Taipa, Macao

[f]: Tenly Inc., Shanghai, China

[g]: Department of Statistics & Data Science, Tsinghua University, Beijing, China

Corresponding Author: Xinshu Zhao

## Abstract

While numerous indices of inter-coder reliability exist, Krippendorff's α and Cohen's κ have long dominated in communication studies and other fields, respectively. The near consensus, however, may be near the end. Recent theoretical and mathematical analyses reveal that these indices assume intentional and maximal random coding, leading to paradoxes and inaccuracies. A controlled experiment with one-way golden standard and Monte Carlo simulations supports these findings, showing that κ and α are poor predictors and approximators of true intercoder reliability. As consensus on a perfect index remains elusive, more authors recommend selecting the best available index for specific situations (BAFS). To make informed choices, researchers, reviewers, and educators need to understand the liberal-conservative hierarchy of indices—i.e., which indices produce higher or lower scores. This study extends previous efforts by expanding the math-based hierarchies to include 23 indices and constructing six additional hierarchies using Monte Carlo simulations. These simulations account for factors like the number of categories and distribution skew. The resulting eight hierarchies display a consistent pattern and reveal a previously undetected paradox in the *Ir* index.

# Liberal-Conservative Hierarchies of Intercoder Reliability Estimators

Measurement validity is a central concern of all disciplines of social sciences based on empirical evidence (Berelson, 1952; Holsti, 1970; Krippendorff, 1980, 2004a, 2012; Neuendorf, 2017; Riffe et al., 1998, 2005, 2014, 2019; Stemler, 2000). *Intercoder reliability,* a.k.a. *interrater reliability,* has been a primary indicator of measurement validity(Hayes & Krippendorff, 2007; Lombard et al., 2002; Lovejoy et al., 2014, 2016; Zhao et al., 2018; Zhao et al., 2013). Researchers also use reliability to evaluate the quality of diagnosis, tests, observations, and other assessments. A widely cited publication in *Biochemia Medica* underscores its importance by directly linking interrater reliability to validity, asserting that it reflects the accuracy of data representation (McHugh, 2012)."

The concept's interdisciplinarity is evident in the variety of terms used across fields, such as inter-annotator reliability and inter-voter reliability, which encompass roles from coder to diagnostician, evaluator to observer. Numerous indices for measuring intercoder reliability have been developed, with Popping (1988) identifying 39 for categorical scales, though many are mathematically equivalent. Zhao et al. analyzed 22 indices, identifying 11 as unique(Zhao, 2011a, 2011b; Zhao et al., 2013). ten Hove et al. (2018) found that 20 indices yielded significantly different results when applied to the same datasets, while Li et al. (2018) used simulations to compare nine indices against each other and generalized coefficients. Out of the dozens, two dominated -- Krippendorff's α for communication research and Cohen's κ for other fields (Cohen, 1960; Hayes & Krippendorff, 2007; Krippendorff, 1970a, 1980; Lombard et al., 2002; Lovejoy et al., 2014, 2016; Zhao et al., 2013). They respectively popularized κ and α. More studies may be expected built in part on the two indices.

The near consensus, however, may begin to end. Since 2013, increasingly more authors complain about the paradoxes and abnormalities regularly produced by the "chance-corrected" indices, especially the “big two,” Cohen’s κ and Krippendorff’s α (Delgado & Tibau, 2019; Dettori & Norvell, 2020; Honda & Ohyama, 2020; Lombard et al., 2004; Nili et al., 2020; ten Hove et al., 2018; Zec et al., 2017; Zhao et al., 2013; Zhao & Zhang, 2014). A controlled experiment of seven most popular indices shows that α, κ and their close relative Scott’s π significantly underperformed the other four indices when predicting or approximating true interrater reliability (Zhao et al., 2022). Notably, the six chance-corrected indices all underperformed *percent agreement* ($a_o$). Given that the chance-corrected indices were designed and declared to outperform percent agreement, question has been raised what legitimate function(s) the indices now serve. Empirical researchers began to report only percent agreement ($a_o$) (Waites et al., 2023), citing the 2022 experiment. More textbooks and methodology reviews now recommend reporting multiple indicators (Riffe et al., 2023; Rojas et al., 2024). A crisis of interrater reliability is looming.

## Crises of Intercoder Reliability and Need for BAFS

Until not too long ago, there was a near consensus across many disciplines that *percent agreement* ($a_o$), the straightforward and widely used index of intercoder reliability, is also the most “primitive” and “flawed,” because it failed to remove chance agreement (Hayes & Krippendorff, 2007). By removing chance agreement, which is the agreement between coders when they code randomly, Cohen’s κ become the most respected index in most disciplines and Krippendorff’s α among communication researchers.

Things began to change after a study published in 2013 (Zhao et al., 2013). The mathematical and behavioral analyses revealed that the chance-corrected indices, including α and κ, assume intentional and maximum random coding by coders (Byrt et al., 1993; Zhao et al., 2018; Zhao et al., 2013). A growing number of reliability experts found this assumption vital and nonsensical (Charles Feng & Zhao, 2016; Checco et al., 2017; Cicchetti & Feinstein, 1990; Feinstein & Cicchetti, 1990; Grove et al., 1981; Krippendorff, 2004b; Scott, 1955). In a published exchange with the authors, Krippendorff delinked intercoder reliability and intercoder agreement, acknowledging that α does not measure intercoder agreement (Krippendorff, 2013), in effect acknowledging that α is not an index of *intercoder reliability* as most of the researchers, including Krippendorff, had defined or understood the term (Riffe et al., 2023; Zhao et al., 2018).

Then came a 2022 controlled experiment that tested seven best-known indices of intercoder reliability against true intercoder reliabilities. The study features two methodological innovations. First a *one-way golden standard* that allows researchers to measure true intercoder reliability as a variable. Second a *reconstructed experiment* method that enables researchers to reorganize individual coding into 384 coding sessions as the subjects of the experiment. The results show that all the six chance-corrected indices tested all significantly underperformed percent agreement, which the six had been designed and declared to outperform, when predicting true reliability. The most respected three, Cohen's κ, Krippendorff's α, and Scott's π, significantly underperformed all the others (Cohen, 1960; Krippendorff, 1970a; Scott, 1955).

Among the most disturbing features of the trio was that each one showed questionable validity, as shown in three aspects --

1) *Chance estimates of κ, α and π are negatively correlated with true chance agreement*. The κ-, α- and π-estimated chance agreements, the unique core of each index, were each negatively correlated with the true chance agreement ($dr^2$=-.152~-.151). That means that the trio tend to remove more agreements when there are fewer chance agreements and remove fewer when there are more. In other words, many “chance agreements” removed by the trio are in fact true agreements, while many “true agreements” not removed by the trio are in fact chance agreements. This feature alone should have invalidated the trio, given that accurately removing chance agreements was the main task of any chance-correcting index.

2) *Indices κ, α and π are poorly correlated with empirical validity, and much more poorly than percent agreement*. In the experiment, the true reliability was highly correlated with correct coding, i.e., empirical validity of the measurement, while each of the trio's estimated reliabilities was poorly correlated with true reliability. Each of the trio also significantly underperformed percent agreement while predicting true reliability ($dr^2$=.841 vs $dr^2$=.312). These findings, when put together, imply that each of the trio is poorly correlated with measurement validity, and each tends to significantly underperform percent agreement when predicting empirical validity. This feature alone should have invalidated the three indices, given that predicting validity is the main reason for estimating reliability and outperforming percent agreement when predicting validity is the main function of any chance-correcting index.

3) *Indices κ, α and π are affected by evenness when they should not be*. The three indices were not only poorly correlated with true reliability ($dr^2$=.312), but they were correlated nearly as much with the evenness of distribution ($dr^2$=.292~.293), which means the three reliability indices measure evenness as much as they measure reliability. The experiment also shows that true reliability was not at all affected by distribution ($dr^2$=.000), which means that reliability indices should not be correlated with evenness at all.

4) *Imposing κ, α* or *π imposes evenness bias, making the world appear flatter*. That *κ, α* and *π* measure evenness has a troublesome implication. Fixed benchmarks, such as $\alpha \geqslant 0.80$ or $\alpha \geqslant 0.667$, are often used to evaluate measurement instruments (Krippendorff, 2004a; Landis, 1977). Since these indices are influenced by distribution evenness, more evenly distributed datasets are more likely to meet these benchmarks, skewing the representation of scientific knowledge towards a "flatter" view of the world, a phenomenon referred to as KAP-imposed evenness bias.

These findings align with a growing body of opinions, analyses, and studies that highlight the pitfalls of α and κ, with some even calling for their banishment (De Vet et al., 2006; Delgado & Tibau, 2019; Feng, 2014; Freelon, 2010; Gwet, 2008; Hoehler, 2000; Jakubauskaite, 2021; Jiang et al., 2021; Kraemer, 1979; Krippendorff, 1970b, 2004b, 2019; Lombard et al., 2002; Riffe et al., 1998, 2005; Stütz et al., 2022; Tong et al., 2020; Xu & Lorber, 2014; Zec et al., 2017; Zhao, 2011a, 2011b; Zhao et al., 2018). This supports the "best available for a situation" (BAFS) approach, which is increasingly advocated by experts. Recognizing that no single index is perfect for all situations, the BAFS approach encourages researchers to select two or more indices that exhibit the least and least harmful deficiencies for the specific research context (Chmura Kraemer et al., 2002; Dettori & Norvell, 2020; Gwet, 2008; Hoehler, 2000; Jiang et al., 2021; Kraemer, 1979, 1992; Li et al., 2018; Nili et al., 2020; ten Hove et al., 2018; Zhao et al., 2018; Zhao et al., 2013).

The BAFS approach requires researchers to know the indices' characteristics, including their tendencies to score high or low, namely their positions on a liberal-conservative scale (Krippendorff, 2019). The knowledge also may help readers when interpreting a research based in part on the scores produced by an intercoder reliability index, such as Cohen's κ (1960) and Krippendorff's α (1980). Since these two seminal publications, numerous studies have cited κ or α to verify or document the empirical basis of their measurement. More studies may be expected built in part on the two indices. Now that the eccentric behavioral assumptions and poor empirical performance of the two indices are better known, systematically investigated liberal-conservative tendencies of the indices may also help in proper interpretation or reinterpretation of the future and past studies.

**Two Functions of Reliability Indices and Need for Hierarchy**

In research practice, intercoder reliability indices perform two functions.

*Function 1, cross-instrument comparison*. One function is to evaluate instruments, e.g., diagnoses, coding, and observations, against each other. The instruments with higher scores are considered more reliable than those with lower scores (Krippendorff, 2016; Riffe et al., 2019; Shrout, 1998).

*Function 2: benchmark comparison*. The second function is to evaluate instruments against fixed benchmarks or benchmark systems. For example, Landis & Koch marked Cohen's κ<0 as poor, κ=0~.2 as slight, .21~.4 as fair, .61~.80 as substantial, and κ >.81 as almost perfect (Cohen, 1960; Silveira & Siqueira, 2023). The system is influential across disciplines (Li et al., 2018). Krippendorff requires α ≥ .8 but also recommends to tentatively accept α ≥ .667 (Krippendorff, 2004a). The system dominates the field of communication (Hayes & Krippendorff, 2007; Shrout, 1998).

For either function, knowledge of the indices' positions on liberal-conservative hierarchies is necessary for proper reading and interpretation of index scores. For cross-instrument comparison, e.g., a Cohen's κ = .81 for one diagnostician may not necessarily indicate higher reliability than a Krippendorff's α = .79 for another diagnostician, if one knows that κ tends to generate higher scores than α with numerous cases of diagnostics and significant differences in prevalence estimations between the diagnosticians (Cohen, 1960; Krippendorff, 1970a, 2004a; Vach, 2005; von Eye & von Eye, 2008; Zhao et al., 2013).

For benchmark comparison, a reader might accept κ=.81 for one study as almost perfect observing the Landis-Koch benchmarks, while another reader might be tentative accepting α=.79 for another study using the Krippendorff benchmarks (Krippendorff, 1980, 2004a, 2012; Silveira & Siqueira, 2023). The readers would be less tempted to do so if they understand that, in this situation, the difference in scores may reflect less different qualities of the instruments, but more the discrepant assumptions of the indices.

Accordingly, this study is tasked to evaluate, cross-verify, and update the collective understanding of the liberal-conservative tendencies of the intercoder reliability indices.

## Liberal and Conservative Tendencies of Intercoder Reliability Indices

The concept of liberal versus conservative scales of reliability indices is not new. Lombard et al (2002) observed that some indices are more “liberal” while others are more “conservative,” which Krippendorff disagreed. Zhao et al. opined that information about numerical patterns can be helpful so long as the information is interpreted with a clear understanding of the concepts and assumptions behind the indices (Krippendorff, 2004a; Zhao et al., 2013).

Based on their analysis of indices’ mathematical formulas, Zhao et al. showed that the numerical values can be dramatically different for different indices, and they produced two liberal-conservative hierarchies for the 22 indices. Similarly, ten Hove et al. found that indices of intercoder reliability generated very different numerical values when applied to the same sets of data (Popping, 1988; ten Hove et al., 2018; Zhao et al., 2013).

However, math-based hierarchies alone may not provide the full picture. Mathematical analyses depend on interpretations of formulas, which can lead to omissions, overemphasis, or errors. This is why modern mathematical studies often pair rigorous derivations with Monte Carlo simulations (Warrens, 2014a). The validity of the hierarchies may be strengthened if they are also informed by empirical data. A main objective of this study is to build additional hierarchies based on simulated empirical data. We hope the two types of hierarchies may verify, complement, correct, and stimulate each other, giving us a more complete picture how the indices behave.

We started with the 22 indices in the two math-based hierarchies built by a social scientist and two mathematical statisticians (Zhao et al., 2013). Being in the existing hierarchies therefore serving as proper references for comparison, the indices also have been analyzed, reanalyzed, tested, retested, debated and re-debated (Byrt et al., 1993; Charles Feng & Zhao, 2016; Chmura Kraemer et al., 2002; Cicchetti & Feinstein, 1990; Delgado & Tibau, 2019; Feinstein & Cicchetti, 1990; Freelon, 2010; Gwet, 2008; Hoehler, 2000; Kraemer, 1979, 1992; Krippendorff, 2004b, 2016, 2019; Shrout, 1998; Vach, 2005; von Eye & von Eye, 2008; Warrens, 2014a, 2014b; Zhao et al., 2018; Zhao et al., 2013; Zwick, 1988).But before we built the simulation-based hierarchies, we identified another index that needs to be added to the math-based hierarchies, which we detail below.

**New Index from Reinterpretation of λr**

Goodman and Kruskal (1959) proposed an agreement index, λr, based on a chance agreement (*ac*) estimation that behaves in some ways similarly to that of Cohen's κ (1960):

$$\boldsymbol{a_c = \frac{1}{2}\left(\frac{N_{l1}}{N} + \frac{N_{l2}}{N}\right)}$$

Some, e.g., Zhao et al.(2013), interpreted $N_{l1}$ and $N_{l2}$ as, respectively, individual frequency reported by each coder, hence (*nl1*+*nl2*)/2, where *nl1*= $N_{l1}/N$ and *nl2*= $N_{l2}/N$, represent the *average frequency* of the two coders. Suppose on a binary scale Coder 1 reports 85 cases in Category 1 and 15 cases in Category 2, while Coder 2 reports 45 cases in Category 1 and 55 cases in Category 2, $N_{l1}$=85, $N_{l2}$=55, and *ac*=(.85+.55)/2=0.7. Goodman and Kruskal's λr shares with major indices the Guttman-Bennet chance-removal formula (Bennett et al., 1954; Guttman, 1946). Inserting Eq. 18 into the classic formula, we have Goodman and Kruskal's λr.

Fleiss (1975), however, interpreted (*nl1*+*nl2*)/2 as the *average frequency* reported by two coders, which in the above example would instead produce an *ac*= (.85+.45)/2=.65. As Goodman and Kruskal did not provide a numerical example, it is not clear which interpretation represents the authors' intention. Thus, we treat the two as two indices, labeling the individual interpretation as λi (Zhao et al., 2013), and the average interpretation as λa (Fleiss, 1975), which makes the total number of indices considered in this study (ten Hove et al., 2018).

## Expanded Math-Based Hierarchies of 23 Indices

In the above example, average interpretation produces a smaller *ac* than individual interpretation. Comparing Table 1 with Table 2, we see it is not a fluke: an *ac* estimation by average interpretation is always smaller than or equal to the counterpart *ac* estimation by individual interpretation, and a smaller *ac* leads to a larger index, λa ≥ λi. The two interpretations differ only when the two coders' estimated distributions are skewed at the opposite directions, e.g., one reports 90/10% while the other reports 5/95%, which is represented by the upper left quarter and the lower right quarter of Table 1 and Table 2. In research practice, these situations are less frequent than the situations represented by the other two quarters, which indicate that the two coders do not disagree with each other in terms of skew directions.

**Table 1**

*Goodman and Kruskal's Chance Agreement ($a_c$) (Individual Interpretation) as a Function of Two Distributions **

| Distribution 2: Positive Findings by Coder 2 ($N_{p2}/N$) in % ** | Distribution 1: Positive Findings by Coder 1 ($N_{p2}/N$) in %** | | | | | | | | | | |
|---|---|---|---|---|---|---|---|---|---|---|---|
| | 0 | 10 | 20 | 30 | 40 | 50 | 60 | 70 | 80 | 90 | 100 |
| 100 | 100.0 | 95.0 | 90.0 | 85.0 | 80.0 | 75.0 | 80.0 | 85.0 | 90.0 | 95.0 | 100.0 |
| 90 | 95.0 | 90.0 | 85.0 | 80.0 | 75.0 | 70.0 | 75.0 | 80.0 | 85.0 | 90.0 | 95.0 |
| 80 | 90.0 | 85.0 | 80.0 | 75.0 | 70.0 | 65.0 | 70.0 | 75.0 | 80.0 | 85.0 | 90.0 |
| 70 | 85.0 | 80.0 | 75.0 | 70.0 | 65.0 | 60.0 | 65.0 | 70.0 | 75.0 | 80.0 | 85.0 |
| 60 | 80.0 | 75.0 | 70.0 | 65.0 | 60.0 | 55.0 | 60.0 | 65.0 | 70.0 | 75.0 | 80.0 |
| 50 | 75.0 | 70.0 | 65.0 | 60.0 | 55.0 | 50.0 | 55.0 | 60.0 | 65.0 | 70.0 | 75.0 |
| 40 | 80.0 | 75.0 | 70.0 | 65.0 | 60.0 | 55.0 | 60.0 | 65.0 | 70.0 | 75.0 | 80.0 |
| 30 | 85.0 | 80.0 | 75.0 | 70.0 | 65.0 | 60.0 | 65.0 | 70.0 | 75.0 | 80.0 | 85.0 |
| 20 | 90.0 | 85.0 | 80.0 | 75.0 | 70.0 | 65.0 | 70.0 | 75.0 | 80.0 | 85.0 | 90.0 |
| 10 | 95.0 | 90.0 | 85.0 | 80.0 | 75.0 | 70.0 | 75.0 | 80.0 | 85.0 | 90.0 | 95.0 |
| 0 | 100.0 | 95.0 | 90.0 | 85.0 | 80.0 | 75.0 | 80.0 | 85.0 | 90.0 | 95.0 | 100.0 |

*: The table was adapted from Zhao et al., (2013). Main cell entries are Goodman and Kruskal's Chance Agreement ($a_c$) in %.

**: $N_{p1}$ is the number of positive answers by Coder *1*, $N_{p2}$ is the number of positive answers by Coder *2*, and $N$ is the total number of cases analyzed

**Table 2**

*Goodman and Kruskal's Chance Agreement ($a_c$) (Average Interpretation) as a Function of Two Distributions**

| | Distribution *1:* Positive Findings by Coder *1* ($N_{p2}/N$) in %** | | | | | | | | | | |
|---|---|---|---|---|---|---|---|---|---|---|---|
| Distribution 2: Positive Findings by Coder 2 ($N_{p2}/N$) in % ** | 0 | 10 | 20 | 30 | 40 | 50 | 60 | 70 | 80 | 90 | 100 |
| 100 | 50.0 | 55.0 | 60.0 | 65.0 | 70.0 | 75.0 | 80.0 | 85.0 | 90.0 | 95.0 | 100.0 |
| 90 | 55.0 | 50.0 | 55.0 | 60.0 | 65.0 | 70.0 | 75.0 | 80.0 | 85.0 | 90.0 | 95.0 |
| 80 | 60.0 | 55.0 | 50.0 | 55.0 | 60.0 | 65.0 | 70.0 | 75.0 | 80.0 | 85.0 | 90.0 |
| 70 | 65.0 | 60.0 | 55.0 | 50.0 | 55.0 | 60.0 | 65.0 | 70.0 | 75.0 | 80.0 | 85.0 |
| 60 | 70.0 | 65.0 | 60.0 | 55.0 | 50.0 | 55.0 | 60.0 | 65.0 | 70.0 | 75.0 | 80.0 |
| 50 | 75.0 | 70.0 | 65.0 | 60.0 | 55.0 | 50.0 | 55.0 | 60.0 | 65.0 | 70.0 | 75.0 |
| 40 | 80.0 | 75.0 | 70.0 | 65.0 | 60.0 | 55.0 | 50.0 | 55.0 | 60.0 | 65.0 | 70.0 |
| 30 | 85.0 | 80.0 | 75.0 | 70.0 | 65.0 | 60.0 | 55.0 | 50.0 | 55.0 | 60.0 | 65.0 |
| 20 | 90.0 | 85.0 | 80.0 | 75.0 | 70.0 | 65.0 | 60.0 | 55.0 | 50.0 | 55.0 | 60.0 |
| 10 | 95.0 | 90.0 | 85.0 | 80.0 | 75.0 | 70.0 | 65.0 | 60.0 | 55.0 | 50.0 | 55.0 |
| 0 | 100.0 | 95.0 | 90.0 | 85.0 | 80.0 | 75.0 | 70.0 | 65.0 | 60.0 | 55.0 | 50.0 |

*: Main cell entries are Goodman and Kruskal's Chance Agreement ($a_c$) in %.

**: $N_{p1}$ is the number of positive answers by Coder *1*, $N_{p2}$ is the number of positive answers by Coder *2*, and $N$ is the total number of cases analyzed.

A cell-by-cell comparison of Table 1 with Table 2 shows that λi's $a_c$ is always larger than or equal to λa's $a_c$, making λi more conservative. Thus, λa is placed above λi in both hierarchies of Table 3. Another comparison between Table 2 and the corresponding table for Scott's π (Liu & Li, 2024)(c.f., Tables 19.3 in Zhao et al., 2013) reveals that λa's chance agreement is always larger than or equal to Scott's, indicating that λa is more conservative than π. Accordingly, λa is placed below π in both hierarchies of Table 3.

These analyses position the two Goodman & Kruskal (1959) indices at the conservative ends of the expanded 23-index hierarchies, with λi at the very bottom in Table 3. Readers may compare Table 1 or Table 2 with corresponding tables in Zhao et al. (2013) to verify that λa and λi are more conservative than the other indices.

This exercise expands the two 22-index hierarchies to make two 23-index hierarchies, which are shown in Table 3. The two hierarchies are *linked*, that is, the relative positions between the two hierarchies can be compared. For example, by placing *Ir* higher in Hierarchy 1 than *β* in Hierarchy 2, the table indicates that *Ir* is more liberal than *β,* although two never appear in the same hierarchy. Accordingly, Hierarchies 1&2 may be seen as two parts of one hierarchy. Table 3 assumes two coders and binary scale and large enough sample. When the number of categories increases to three and beyond, *S, Ir,* their equivalents, and *AC1* can become more liberal; when a sample reduces to 20 or below, Krippendorff's α can become very liberal (Zhao et al., 2018; Zhao et al., 2013).

**Table 3**

*Two Liberal-Conservative Hierarchies Based on Mathematical Analyses of Reliability Indices* (modification from Zhao et al., 2013)*

| | Hierarchy 1 | Hierarchy 2 |
|---|---|---|
| More *liberal* estimates of reliability. | Percent Agreement ($a_o$) (pre 1901), Osgood's (1959) index, Holsti's *CR* (1969), Rogot & Goldberg's $A_1$ (1966) | Percent Agreement ($a_o$) (pre 1901), Osgood's (1959) index, Holsti's *CR* (1969) Rogot & Goldberg's $A_1$ (1966) |
| ↑ | Perreault & Leigh's $I_r$ (1989) | |
| | Gwet's $AC_1$ (2008, 2010) | |
| | Guttman's $\rho$ (1946), Bennett et al.'s *S* (1954), Guilford's *G* (1961), Maxwell's *RE* (1977), Jason & Vegelius' *C* (1979), Brennan & Prediger's $k_n$ (1981), Byrt et al.'s *PABAK* (1993) Potter & Levine-Donnerstein's *rdf-Pi* (1999). | |
| | | Cohen's κ (1960) Rogot & Goldberg's $A_2$ (1966) |
| | Krippendorff's α (1970, 1980) | Krippendorff's α (1970, 1980) |
| ↓ | Scott's π (1955), Siegel & Castellan's *Rev-K* (1988), Byrt et al's BAK (1993) | Scott's π (1955), Siegel & Castellan's *Rev-K* (1988), Byrt et al's BAK (1993) |
| | Goodman & Kruskal's $\lambda_a$ (1954) | Goodman & Kruskal's $\lambda_a$ (1954) |
| More *conservative* estimates of reliability. | Goodman & Kruskal's $\lambda_i$ (1954) | Goodman & Kruskal's $\lambda_i$ (1954) |

* The two hierarchies assume binary scale, two coders, and sufficiently large sample. Comparisons across the dotted lines are between the general patterns in situations that are more frequent and more important for typical research, e.g., when indices are zero or above, and when the distribution estimates of two coders are not extremely skewed in opposite directions. Comparisons involving Guttman's $\rho$, its eight equivalents, and Perreault & Leigh's $I_r$ assume binary scale. Comparisons involving Krippendorff's α assume sufficiently large sample.

Table 3 assumes two coders and binary scale and large enough sample. When the number of categories increases to three and beyond, *S, Ir,* their equivalents, and *AC1* can become more liberal; when a sample reduces to 20 or below, Krippendorff's α can become very liberal (Zhao et al., 2018; Zhao et al., 2013).

**Six Simulation-Based Hierarchies**

The two expanded hierarchies above are based on mathematical analysis. Two cells are separated by a solid line only if the index(es) in an upper cell is usually larger and never smaller than the index(es) in the lower cell. When one index is sometimes larger and sometimes smaller than the other, a judgment call was made whether to put them in a same cell, in two cells separated by dotted lines, or into different hierarchies.

A data-based hierarchy may help reduce uncertainty by showing whether one index consistently yields higher scores than another, thus identifying it as more liberal. Additionally, mathematical analysis is more efficient with simpler situations and is limited to binary scales. As complexity increases, such as when the number of categories increases from two to three, the mathematical analysis becomes exponentially more complex and prone to errors. A data-based approach is more efficient for handling these more complicated situations.

Accordingly, a Monte Carlo simulation was performed to build a liberal-conservative hierarchy based on data. We manipulated two between-subjects factors, i.e., the number of categories (three levels, i.e., 2, 5 and 9 categories) and sample sizes (12 levels, i.e., 10, 13, 14, 15, 16, 17, 20, 25, 30, 100, 500, and 2,000). Each condition has 5,000 contingency tables, so the total sample size is 180,000. Eight intercoder reliability indices were derived from each contingency table.

Multiple comparisons with Tukey's HSD were conducted to examine the mean differences among the indices. This procedure allows us to detect which index yields higher values and which gives lower values. In Table 4, an index is placed in a higher cell than another when the former produces a higher mean value than the latter and the difference is statistically acknowledged, aka significant, at $p<0.05$. Indices are placed in the same cell when the difference between their mean values is statistically non-acknowledged.

**Table 4**

*Six Liberal-Conservative Hierarchies Based on Monte Carlo Simulation*

| | *Hierarchy 3* | *Hierarchy 4* | *Hierarchy 5* | *Hierarchy 6* | *Hierarchy 7* | *Hierarchy 8* |
|---|---|---|---|---|---|---|
| | $C$=2; $N$ = 2,000 Distribution restricted to 45%~55% | $C$=2; $N$ = 2,000; un-restricted; distribution | $C$=2; $N$ = 12 levels; un-restricted; distribution | $C$=5; $N$ = 12 levels; un-restricted; distribution | $C$=9; $N$ = 12 levels; un-restricted; distribution | $C$ = 2, 5 & 9; $N$ = 12 levels; un-restricted; distribution |
| *Most Liberal* | $a_o$ | $a_o$ | $a_o$ | $a_o$ | $I_r$ | $a_o$ |
| | $I_r$ | $I_r$ | $I_r$ | $I_r$ | $a_o$ | $I_r$ |
| | | $\kappa$ | $\kappa$ | | | $\kappa$ |
| | | $AC_1$ | $AC_1$ | $\kappa$, $AC_1$ | $\kappa$, $AC_1$, $S$ | $AC_1$ |
| | | $S$ | $S$ | $S$ | | $S$ |
| | | | α | | | α |
| | π, α, κ, $AC_1$, $S$ | π, α | π | π, α | π, α | π |
| *Most Conservative* | $\lambda_a$ | $\lambda_a$ | $\lambda_a$ | $\lambda_a$ | $\lambda_a$ | $\lambda_a$ |
| | $\lambda_i$ | $\lambda_i$ | $\lambda_i$ | $\lambda_i$ | $\lambda_i$ | $\lambda_i$ |

When the analysis is based on the entire data, which includes all three levels of categories (*K*=2, 5&9*)*, 12 sample sizes (*N*=12 levels), and unrestricted distribution, percent agreement (*a*o) is the most liberal, followed by *Ir*, κ, *AC1* and *S*. Goodman and Kruskal's λi is consistently the most conservative. Krippendorff's α and Scott's π are in between.

With nine or five categories, a similar pattern emerged. With nine categories, the differences among κ, *AC1*, and *S*, and between π and α become statistically non-acknowledged. In addition, *Ir* is even more liberal than percent agreement, which we will discuss further in another section. With five categories, the differences between κ and *AC1*, and between π and α become not acknowledged.

With two categories and 2,000 cases, the only difference that is statistically non-acknowledged is between π and α. Since some indices like κ are very sensitive to the skewness of marginal distributions, we ran another test after restricting distribution to within .45~.55. Under this condition, the differences between κ, *AC1*, π, α and *S* become statistically non-acknowledged. This shows that these five indices are very similar to each other with two categories, moderately even distributions, and sufficiently large samples.

At the "starting line"—two categories, an infinitely large sample, and a 50-50% distribution—these indices are equal. As the number of categories increases while other factors remain unchanged, S and AC1 increase rapidly, while the other indices lag. When the distribution becomes more skewed from the starting line, π, κ, and α decrease, AC1 increases, and S remains stable. When the sample size decreases from the starting line, α increases while other indices remain unchanged.

In general, percent agreement is the most liberal, while λi is the most conservative. This is because percent agreement assumes no chance agreement, yet the chance agreement of λi always chooses the largest marginal distribution. The values of α are higher than those of π, lower than those of *S*. The values of *AC1* are usually between κ and *S*. *Ir* is the second most liberal index next to percent agreement. *Ir*, *S* and *AC1* are influenced by the number of categories, so their reliability values will vary with the change of number of categories. The difference between π and α is neglegible when sample siezes get very large. Since κ, π and α are dependent on distribution skews, they become indistinguishable from *S* and *AC1* when marginal distributions get even.

**Non-adjusted, Category-based, and Distribution-based Indices**

Comparing the hierarchies in Table 3 and Table 4, a pattern emerges: non-adjusted indices like percent agreement (ao) are the most liberal, distribution-based indices are the most conservative, and category-based indices are in between. Gwet's AC1, a double-based index, is closer to category-based indices. This pattern aligns with the underlying assumptions of the indices. Non-adjusted indices assume no chance agreement, making them the most liberal. Distribution-based indices assume that skewed distributions increase chance agreement, leading to more conservative values. Category-based indices assume that more categories reduce chance agreement, even if empty, making them more liberal than distribution-based indices.

As expected, there are differences between the mathematics-based and the simulation-based hierarchies. Most notably, Cohen's κ occupies more liberal positions in the simulation-based Hierarchies 3~8 than in the mathematics-based Hierarchies 1 & 2. This is due to κ's unique individual quota assumption, which expects low chance agreement with "contrasting skews" (e.g., one coder reports 80% positive while the other reports 80% negative). In extreme cases, such as one coder reporting 100% positive and the other 0% positive, κ expects no chance agreement removal, leading to higher κ values. In our Monte Carlo simulation, data are generated randomly, resulting in an equal number of "contrasting skews" and "congruent skews" (where both coders' distributions are skewed in the same direction). In their mathematical analysis, Zhao et al. (2013) placed less weight on contrasting skews, which are less common in typical research, leading to a more conservative placement of κ in the mathematics-based hierarchies.

Based on mathematical analysis, α is listed as more liberal than π in both hierarchies in Table 3. Based on simulated data, α is also listed as more liberal than π in two of the five hierarchies in Table 4. The difference, however, become statistically non-acknowledged in the other four hierarchies. The phenomenon is not surprising. In observed or simulated data, as sample size (compare Hierarchies 5 with 3 or 4) or categories (compare Hierarchies 8 with 6 or 7) increase, the differences between the two indices can become so small that they are statistically non-acknowledged. But in mathematical analysis, a tiny difference is still a difference, so α is still listed as more liberal than π.

**A New Paradox for Perreault & Leigh's *Ir***

Perreault & Leigh's *Ir* (1989) produced an interesting standout in Hierarchy 4, where *Ir* appears even more liberal than percent agreement *ao*. *Ir* was designed to adjust for, which means to remove, chance agreement. Removing chance agreement is not supposed to make a reliability index larger, hence a new paradox to be added to the 22nd paradox listed by Zhao et al. (2013):

***Paradox 23:*** Reliability index appears larger after removing random chance agreement.

The paradox is due to the combined effects of three assumptions behind Ir, maximum randomness, categories reduce chance agreements, and index needs to be elevated (Zhao et al., 2013).

In the traditional approach, the maximum randomness assumption acts as a double-edged sword. While it suppresses the index by subtracting the maximum chance agreement (*ac*) from the numerator in Equation 3, it also inflates the index by subtracting the same ac from the denominator, shortening the reference scale. On its own, this assumption would not result in a chance-adjusted index larger than *ao*.

However, *Ir* is also category-based, and its chance estimation reduces quickly as the number of categories increases, even when additional categories are unused, which further inflates the index. Yet, this alone does not make *Ir* surpass *ao*. The key factor is the assumption that the index needs to be elevated, achieved in *Ir* by taking the square root of S, which ultimately allows *Ir* to exceed *ao*.

Although in Table 4 the paradox appears only in Hierarchy 7, where there are nine categories, the paradox can happen in many other situations, and the underlying problem is more pervasive. Our simulation skipped $K$=6 through $K$=8 and stopped at $K$=9. The paradox did not show up in Hierarchies 6 ($K$=5) and 8 ($K$=2, 5 & 9) not because *Ir* never passed *ao*, but because *Ir* did not pass *ao* far enough or often enough to make the average *Ir* larger than average *ao*. But even at $K$=4 or $K$=3, *Ir* can be larger than *ao*. To verify, set $K$=3, and *ao* any number larger than 0.5 but smaller than 1, and calculate $Ir = ((1- 1/K) / (ao-1/K))^2$. This analysis confirms that the line between *ao* and *Ir* in Table 12 of Zhao et al. (2013) should be dotted as it is. In other words, while *ao* is more often larger than *Ir*, *Ir* is also very often larger than *ao*.

This is not just about one index slightly larger or smaller than another. Zhao et al. recommended not to use *Ir* or *S* when there are three or more categories (Zhao et al., 2013). The newly discovered paradox and behavior of *Ir* seem to suggest not to use *Ir* even with two categories, where *Ir* never exceeds *ao*, *Ir* may be overly inflated. *Ir* has no advantage over *S* under any circumstance, and in almost all circumstances *S* is a more reasonable alternative to *Ir*, the drawbacks of *S* notwithstanding. The only exceptions are when *S*=0 or *S*=1, where *S*= *Ir*,

## Conclusion

This study extended Zhao et al.'s work, using simulated data to build six more hierarchies which show many similarities with previous hierarchies but also show differences (Zhao et al., 2013). Some preliminary findings of this study were reported in a 2012 conference presentation (Zhao et al., 2012). We hope these hierarchies together will prove useful to workbench researchers who wish to better evaluate the inter-coder reliability indices of their instruments.

Findings from the mathematical analysis and the simulation are consistent with each other on some main points. Between groups of the indices, the non-adjusted indices tend to be the most liberal, the distribution-based indices tend to be the most conservative, and the category-based indices tend to be somewhere in between. Between the individual indices, percent agreement tends to be the most liberal, while Goodman and Kruskal's $\lambda r$ tends to be the most conservative. Between the distribution-based indices, $\kappa$ tends to be more liberal than $\alpha$, which tends to be more liberal than $\pi$, which tends to be more liberal than $\lambda r$.

Three discrepancies emerged. First, $\kappa$ appears more liberal in simulation than in mathematical analysis. Second, $\alpha$ appears more liberal than $\pi$ in mathematical analysis while the two appeared tied statistically in the simulation analysis. Third, *Ir* appears more liberal in simulation than in mathematical analysis. Our analysis shows that mathematical analysis is more precise in the first two discrepancies, while in the third discrepancy simulation filled a gap in the mathematical analysis.

Researchers want their reliabilities to look high, and they have many indices to choose from. Two newer indices, Perreault & Leigh's *Ir* and Gwet's *AC1*, are gaining in popularity in part because they tend to produce higher numbers than other indices. Knowing that they are among the most liberal, we hope, would encourage the researchers, reviewers and editors to be more cautious when using or interpreting the two indices. On the other hand, we should not equate low estimate with rigor, or complex formulas with sophistication. More specifically, reviewers should not require $\lambda r$, $\pi$ or $\alpha$ just for its low estimates or complicated equations. We also should not require or encourage universal application of $\alpha$ just because it has been repeatedly advocated.

What we very much need is a criterion or criteria that will help us to evaluate which index is more appropriate or accurate for various research situations. Ultimately, we need a new index(es) based more realist assumptions of coder behavior. These assumptions should include 1) coders sometimes code randomly, which leads to random agreements that need to be removed; 2) the random coding is not deliberate or purposeful, therefore chance agreement is not a function of category or distribution; 3) the random coding is involuntary depending on difficulty of the task; a more difficult task produces more chance agreement, a less difficult task produces less chance agreement, a non-difficult task produces no chance agreement.

## Declarations

### Ethics approval and consent to participate

The experiment mentioned in this research received ethical approval under the ethics procedures of University of Macau Panel on Research Ethics (reference SSHRE22-APP016-FSS). Written consent for the survey was also taken.

All methods were carried out in accordance with relevant guidelines and regulations. Informed consent was obtained from all subjects and/or their legal guardian(s).

### Consent for publication

Not applicable.

### Availability of data and materials

The datasets used and/or analyzed during the current study are available from the corresponding author on reasonable request.

### Competing interests

The authors declare that they have no competing interests.

### Funding

This research is supported in part by grants of University of Macau, including CRG2021-00002-ICI, ICI-RTO-0010-2021, CPG2022-00004-FSS and MYRG2020-00233-FSS, ZXS PI; Macau Higher Education Fund, HSS-UMAC-2020-02, ZXS PI; Jiangxi 2K Initiative through Jiangxi Normal University School of Journalism and Communication, 2018-08-10, Zhao PI.

### Acknowledgements

The authors gratefully acknowledge the contributions of Hui Huang to the execution of the reconstructed experiment.

### Authors' contributions

XZ drafted the overall design and the theoretical framework, to which HH contributed revision. XZ also designed and executed the mathematical analyses and drafted the initial manuscript. GCF and XZ designed the simulations, which GCF executed and YC replicated. YJZ and DML contributed to the redesign of the study, the replication and revision of the mathematical analysis, and the manuscript rewriting. SHA, MML, and ZTT contributed to the reviews of literature, replication of mathematical analyses, revision of the theories, and numerous rewritings of the manuscript. All authors read and approved the final manuscript.